\documentstyle[sprocl,psfig]{article}

\def\be{\begin{equation}}
\def\ee{\end{equation}}
\def\bea{\begin{eqnarray}}
\def\eea{\end{eqnarray}}
\begin{document}

\title{REALISTIC EFFECTIVE INTERACTIONS AND LARGE--SCALE NUCLEAR STRUCTURE
CALCULATION}

\author{T.~Engeland, A.~Holt  and E.~ Osnes}

\address{Department of Physics, University of Oslo,\\
         Postbox  1048, 0316 Oslo, Norway}

\author{M.~Hjorth-Jensen}

\address{Nordita, Blegdamsvej 17,\\
                  DK-2100 K\o benhavn \O, Denmark}

%%%%%%%%%%%%%%%%%%%%%%%%%%%%%%%%%%%%%%%%%%%%%%%%%%%%%%%%%%%%%%
% You may repeat \author \address as often as necessary      %
%%%%%%%%%%%%%%%%%%%%%%%%%%%%%%%%%%%%%%%%%%%%%%%%%%%%%%%%%%%%%%

\maketitle\abstracts{We describe the properties of complex nuclei,
 such as the Sn isotopes with mass numbers A = 100 -- 132,
in terms of the free nucleon--nucleon
 interaction as obtained from meson--exchange theory. This amounts to first calculating
 an effective interaction in which the  free interaction is modified by the 
 presence of the appropriate nuclear medium. The short--range correlations are
 included within the framework of Brueckner theory yielding the nuclear reaction
 matrix and the long--range correlations by using the reaction matrix in many--body
 perturbation theory to obtain an effective interaction. The resulting effective 
 interaction is then employed in calculating the nuclear properties.
Particular emphasis is placed on
 the ability of our calculation to describe systematic trends of the properties
 of these nuclei. Both successful achievements and problematic features are 
pointed out.}

\section{Introduction}

One of the fundamental, yet unsolved problems of nuclear theory is to describe the properties of
 complex nuclei in terms of their constituent particles and the interaction among them.
 There are two major obstacles to the solution of this problem. Firstly, we are dealing
 with a quantal many--body problem which cannot be solved exactly. Secondly, the basic
 nucleon--nucleon (NN) interaction is not well known. Thus, it may be difficult to know 
whether an eventual failure to solve this problem should be ascribed to the many--body 
methods or the interaction model used. On the other hand, these two uncertainties are 
intimately connected. In any model chosen to approximate the original many--body problem
 one has to apply an interaction which is consistent with the particular degrees of freedom 
considered. This amounts to correcting the original interaction for the degrees of freedom 
not explicitly included in the many--body treatment, thus yielding a so--called effective interaction.

In principle, one should start from an NN interaction derived from the interaction between quarks. 
Although attempted, this program has not been quantitatively successful. Thus, one has to be 
content with using an NN interaction derived from meson--exchange models which reproduce the 
relevant two--nucleon data. Such interactions are generally termed realistic interactions. 
Examples are the Paris, Bonn and Nijmegen potentials.

Once the basic NN interaction has been established, it should be employed in a quantal 
many--body approximation to the nuclear structure problem of interest. One such approximation 
is the spherical shell model, which has provided a successful microscopic approach for nuclei 
near closed shell. Away from closed shells the number of valence particles and single--particle 
orbits quickly becomes too large for the shell model to handle. On the other hand, there has 
been enormous progress in computer technology over the past years and this trend is likely 
to continue. It is therefore a challenge to use modern hardware computer technology coupled 
to effecient numerical methods developed in other fields of science and technology to handle  complex 
nuclear structure problems.

This would indeed allow us to test the theory of realistic effective interactions 
in nuclei with many valence nucleons. Hitherto, realistic nuclear forces have 
mainly been applied to nuclei with two or a few valence particles beyond 
closed shells, such as the oxygen and calcium isotopes. Thus, by going to the 
tin isotopes, in which the major neutron shell between neutron numbers 50 and 
82 is being filled beyond the $^{100}$Sn closed shell core, we have the opportunity 
of testing the potential of large--scale shell model calculations as well as the realiability of 
realistic effective interactions in systems with many valence particles. It should 
be noted that in many current shell model calculations the effective interaction 
is frequently either parametrized or adjusted in order to optimize the fit to the 
data. As a matter of principle we shall refrain from making any such 
adjustments and stick to the interaction obtained by a rigorous calculation 
consistent with the many--body scheme chosen. Only then one will be able to 
assess the quality and reliability of the interaction obtained and the possible 
needs for improvement. One limitation of the present work is that we only 
consider effective two--body forces. In systems with many valence particles one 
should in principle also include effective three- and many--body forces. This 
will however have to be deferred to future work.

In addition to being an exploratory calculation testing the shell model and 
realistic effective interactions over a wide range of Sn isotopes, the present 
work is also a calculation in its own right. Both the low--mass, very 
neutron--deficient and the heavy--mass, neutron rich Sn isotopes are very unstable and 
have only recently been identified and become accessible to spectroscopic 
studies. Thus, they represent a challenge to theoretical work as well. 

\newpage
The paper is organized as follows. In Sect.~2 we give a brief summary of the 
calculation of the effective interaction. Then, in Sect.~3 we discuss our 
shell model algorithm. The results are presented in Sect.~4 and concluding remarks in 
Sect.~5.

\section{Calculation og the shellmodel effective interaction}
The aim of microscopic nuclear structure calculations is to derive
various properties of finite nuclei from the underlying 
Hamiltonian describing the interaction between 
nucleons. 
When dealing with nuclei, such as the tin isotopes with $A\sim 100$, 
the full dimensionality of the 
many--body Schr\"{o}dinger equation for an $A$--nucleon system
\begin{equation}
     H\Psi_i(1,...,A)=E_i\Psi_1(1,...,A),
     \label{eq:full_a}
\end{equation}
becomes intractable and one has to seek 
viable approximations to Eq.(\ref{eq:full_a}). 
In Eq.(\ref{eq:full_a}), $E_i$ and $\Psi_i$ 
are the eigenvalues and eigenfunctions
for a state $i$ in the Hilbert space.

In nuclear structure calculations, one is normally 
only interested in solving Eq.(\ref{eq:full_a})
for certain
low--lying states. It is then customary to divide the Hilbert space
into a model space defined by the operator $P$ and an excluded space
defined by the operator $Q$
\begin{equation}
     P=\sum_{i=1}^{d}\left | \psi_i\right\rangle 
     \left\langle\psi_i\right |
     \mbox{\hspace*{1cm}}
    Q=\sum_{i=d+1}^{\infty}\left | \psi_i\right\rangle 
     \left\langle\psi_i\right | ,
\end{equation}
with $d$ being the size of the model space and such that $PQ=0$.
The assumption then is that the components of these low--lying
states can be fairly well reproduced by configurations consisting
of a few particles/holes occupying physically selected orbits.
These selected orbitals define the model space.

Eq.(\ref{eq:full_a})  can then be rewritten as a secular equation
\begin{equation}
    PH_{\mathrm{eff}}P\Psi_i=P(H_{0}+V_{\mathrm{eff}})
    P\Psi_i=E_iP\Psi_i,
\end{equation}
where $H_{\mathrm{eff}}$  now is an effective Hamiltonian acting solely
within the chosen model space. The term $H_0$
is the unperturbed Hamiltonian while the effective interaction is
given by
\begin{equation}
  V_{\mathrm{eff}}=\sum_{i=1}^{\infty} V_{\mathrm{eff}}^{(i)},
\end{equation}
with $ V_{\mathrm{eff}}^{(1)}$,  $ V_{\mathrm{eff}}^{(2)}$,
 $ V_{\mathrm{eff}}^{(3)}$,...\ being effective one--body, two--body,
three--body interactions etc. 
It is also customary in nuclear shell model calculations to add
the one--body effective interaction  $ V_{\mathrm{eff}}^{(1)}$
to the unperturbed part of the Hamiltonian so that
\begin{equation}
    H_{\mathrm{eff}}= \widetilde{H}_{0}+  V_{\mathrm{eff}}^{(2)}+
    V_{\mathrm{eff}}^{(3)}+\dots,
\end{equation}     
where $\widetilde{H}_{0}=H_{0}+V_{\mathrm{eff}}^{(1)}$. This allows us,
as in the shell model, to replace the eigenvalues of 
$\widetilde{H}_{0}$ by the empirical single--particle energies 
for the nucleon orbitals of our model space, or valence space, e.g..,
$2s_{1/2}$, $1d_{5/2}$, $1d_{3/2}$, $0g_{7/2}$ and $0h_{11/2}$ for
Sn isotopes under consideration,
Thus, the remaining quantity to calculate is the two- or more--body
effective interaction 
$\sum_{i=2}^{\infty} V_{\mathrm{eff}}^{(i)}$.
In this work we will restrict our attention to the derivation of
an effective two--body interaction 
\begin{equation}
      V_{\mathrm{eff}}=V_{\mathrm{eff}}^{(2)},
\end{equation}
using the many--body methods discussed in \mbox{Ref. \cite{hko95}}
and reviewed below.
The study of effective three--body forces will be deferred to a later
\mbox{work \cite{eh97}}. 

Our scheme to obtain an effective two--body interaction for 
the tin isotopes
starts with a free nucleon--nucleon  interaction $V$ which is
appropriate for nuclear physics at low and intermediate energies. 
At present, there are several potentials available. The most recent 
versions of Machleidt and \mbox{co--workers \cite{cdbonn}},
 the Nimjegen \mbox{group \cite{nim}} and the Argonne
\mbox{group \cite{v18}} have a $\chi^2$ per datum close to $1$.
In this work we will thus choose to work with the charge--dependent
version of the Bonn potential models, see \mbox{Ref. \cite{cdbonn}}.
The potential model of \mbox{Ref. \cite{cdbonn}} is an extension of the 
one--boson--exchange models of the Bonn \mbox{group \cite{mac89}}, where mesons 
like $\pi$, $\rho$, $\eta$, $\delta$, $\omega$ and the fictitious
$\sigma$ meson are included. In the charge--dependent version
of \mbox{Ref. \cite{cdbonn}}, the first five mesons have the same set
of parameters for all partial waves, whereas the parameters of
the $\sigma$ meson are allowed to vary. 

The next step 
in our perturbative many--body scheme is to handle 
the fact that the repulsive core of the nucleon--nucleon potential $V$
is unsuitable for perturbative approaches. This problem is overcome
by introducing the reaction matrix $G$ given by the solution of the
Bethe--Goldstone equation
\begin{equation}
    G=V+V\frac{Q}{\omega - H_0}G,
\end{equation}
where $\omega$ is the unperturbed energy of the interacting nucleons,
and $H_0$ is the unperturbed Hamiltonian. 
The operator $Q$, commonly referred to
as the Pauli operator, is a projection operator which prevents the
interacting nucleons from scattering into states occupied by other nucleons.
In diagrammatic language 
the $G$--matrix is the sum over all
ladder type of diagrams. This sum is meant to renormalize
the repulsive short--range part of the interaction. The physical interpretation
is that the particles must interact with each other an infinite number
of times in order to produce a finite interaction. 

Since the perturbative interaction is $V - U$ rather than $V$, $U$ beeing
an auxiliary one--body potential incorporated in $H_0$ along with the kinetic 
energy $T$, it is convenient to include insertions of $U$ to arbitrary order
in the intermediate states of $G$. This can be done by redefining $G$ as 
\[
G = V + V\frac{Q}{\omega - QTQ}G.
\]
It is further convenient to calculate $G$ using the double--partioning scheme discussed
in e.g.,~\mbox{Ref. \cite{hko95}}.
A harmonic--oscillator basis was chosen for the
single--particle
wave functions, with an oscillator energy $\hbar\Omega$ given
by
$\hbar\Omega = 45A^{-1/3} - 25A^{-2/3}=7.87 $ MeV,  $A$ being the mass
number.

\begin{center}
	\begin{figure}[hbtp]
	\begin{center}
	\psfig{figure=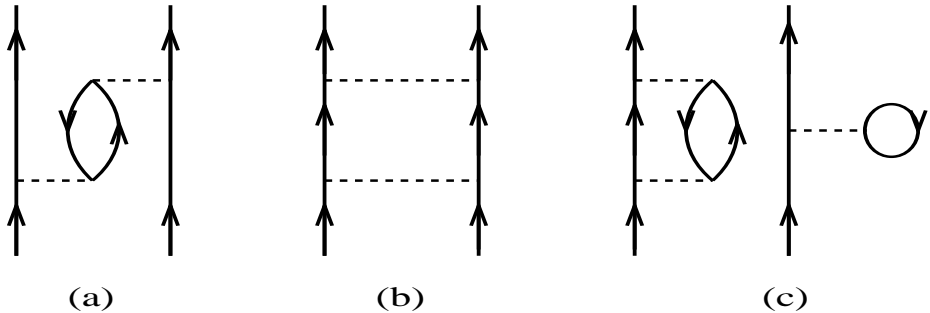,width=12cm,height=4cm}
	\end{center}
	\caption{Different types of valence--linked diagrams. Diagram (a)
	is irreducible and connected, (b) is reducible, while (c) is irreducible
	and disconnected.}
	\label{fig:fig1}
	\end{figure}
	\end{center}

Finally, we briefly sketch how to calculate an effective 
two--body interaction for the chosen model space
in terms of the $G$--matrix.  Since the $G$--matrix represents just
the summation to all orders of particle--particle
ladder diagrams, there are obviously other terms which need to be included
in an effective interaction. Long--range effects represented by 
core--polarization terms are also needed.
The first step then is to define the so--called $\hat{Q}$--box given by
\begin{equation}
   P\hat{Q}P=PGP +
   P\left(G\frac{Q}{\omega-H_{0}}G\\ + G
   \frac{Q}{\omega-H_{0}}G \frac{Q}{\omega-H_{0}}G +\dots\right)P.
   \label{eq:qbox}
\end{equation}
The $\hat{Q}$--box is made up of non--folded diagrams which are irreducible
and valence linked. A diagram is said to be irreducible if between each pair
of vertices there is at least one hole state or a particle state outside
the model space. In a valence--linked diagram the interactions are linked
(via fermion lines) to at least one valence line. Note that a valence--linked
diagram can be either connected (consisting of a single piece) or
disconnected. In the final expansion including folded diagrams as well, the
disconnected diagrams are found to cancel \mbox{out \cite{ko90}}.
This corresponds to the cancellation of unlinked diagrams
of the Goldstone \mbox{expansion \cite{ko90}}. 
We illustrate these definitions by the diagram shown
in Fig. \ref{fig:fig1}, where an arrow pointed upwards(downwards) is a particle(hole)
state. 
Diagram (a) is irreducible, valence linked and connected,
while (b) is reducible since the intermediate particle states belong to the model space
(particle states outside the model space would be denoted by railed lines). Diagram (c)
is reducible, valence linked and disconnected.

\newpage

We can then obtain an effective interaction
$H_{\mathrm{eff}}=\widetilde{H}_0+V_{\mathrm{eff}}^{(2)}$ in terms of the $\hat{Q}$--box,
 using the folded--diagam \mbox{expansion \cite{hko95,ko90}}
\begin{equation}
    V_{\mathrm{eff}}^{(2)}(n)=\hat{Q}+{\displaystyle\sum_{m=1}^{\infty}}
    \frac{1}{m!}\frac{d^m\hat{Q}}{d\omega^m}\left\{
    V_{\mathrm{eff}}^{(2)}(n-1)\right\}^m,
    \label{eq:fd}
\end{equation}
where $(n)$ and $(n-1)$ refer to the effective interaction after
$n$ and $n-1$ iterations. The zeroth iteration is represented by just the 
$\hat{Q}$--box.
Observe also that the
effective interaction $V_{\mathrm{eff}}^{(2)}(n)$
is evaluated at a given model space energy
$\omega$, as is the case for the $G$--matrix as well. Here we choose
$\omega =-20$ MeV.
Less than $10$ iterations were needed in order to obtain a numerically
stable result. Note that all non--folded diagrams through 
third--order in the interaction $G$ are included in the $\hat{Q}$--box.
For further details, see \mbox{Ref. \cite{hko95}}.

Another iterative scheme which has been much favored in the literature is one
proposed by Lee and \mbox{Suzuki \cite{leeSu}}. However, contrary to the folded--diagram
expansion of Eq.(\ref{eq:fd}), which gives those states having the largest overlap 
with the model space states, the Lee--Suzuki method  converges to the lowest eigenstates
regardless of their overlaps with the model space.
Thus, Eq.(\ref{eq:fd}) seems more
 \mbox{appropriate \cite{paul}} for shell model calculations than the Lee--Suzuki scheme.

\section{The shell model scheme}
The effective two--particle interaction can in turn be used in shell model
calculations.
Both binding energies and excitation spectra are severe tests of the method.
Furthermore, it is of importance to analyze nuclear systems with large number of degrees
of freeedom. Thus in the present work we have chosen the Sn region with 
$Z = 50$ and $50 < N  <82$ and limit ourselves to effective two--particle
matrix elements with $T = 1$. Two types of calculation are performed:
\vspace*{10pt}

\begin{minipage}{0.9\textwidth}
(I). Effective two--particle matrix elements are calculated based on 
a $Z = 50, N = 50$ symmetric core and with the active $P$--space
based on the single--particle orbits 
$2s_{1/2}$, $1d_{5/2}$, $1d_{3/2}$, $0g_{7/2}$ 
and $0h_{11/2}$.
The corresponding single--particle energies are not known experimentally.
At present we choose $\varepsilon(d_{5/2}^{+}) = 0.00$~MeV, 
$\varepsilon(g_{7/2}^{+}) = 0.08$~MeV,
$\varepsilon(s_{1/2}^{+}) = 2.45$~MeV, $\varepsilon(d_{3/2}^{+}) = 2.55$~MeV
and $\varepsilon(h_{11/2}^{-}) = 3.20$~MeV. 
These data are in reasonable agreement with 
similar shell model calculations in this region, see for example \mbox{Ref. \cite{aldo}}.
However, the single--particle energies of $s_{1/2}$ and  $d_{3/2}$ have been adjusted 
in order to 
reproduce the lowest $1/2^{+}$ and $3/2^{+}$ states in $^{111}$Sn, see the discussion in 
\mbox{Ref. \cite{nico}}. The shell model calculation then amounts to studying
valence neutrons outside this core.
\vspace*{10pt}

(II). Effective two--hole matrix elements are calculated based on 
a $Z = 50, \quad N = 82$ asymmetric core and with the active $P$--space for holes
based on the $2s_{1/2}$, $1d_{5/2}$, $1d_{3/2}$, $0g_{7/2}$ and $0h_{11/2}$
hole orbits.
The corresponding single--hole energies
$\varepsilon(d_{3/2}^{+}) = 0.00$~MeV, 
 $\varepsilon(h_{11/2}^{-}) = 0.242$~MeV, $\varepsilon(s_{1/2}^{+}) = 0.332$~MeV,
$\varepsilon(d_{5/2}^{+}) = 1.655$~MeV and  $\varepsilon(g_{7/2}^{+}) = 2.434$~MeV
are taken from \mbox{Ref. \cite{jan}}
and the shell model calculation amounts to studying
valence neutron holes outside this core.
\end{minipage}
\vspace*{10pt}

The shell model problem requires the solution of a real symmetric
$n \times n$ matrix eigenvalue equation
\begin{equation}
       \widetilde{H}\left | \Psi_k\right\rangle  = 
       E_k \left | \Psi_k\right\rangle .
       \label{eq:shell_model}
\end{equation}
where for the present cases the dimension of the $P$--space reaches $n \approx 2 \times 10^{7}$.
At present our basic approach in finding solutions to Eq.(\ref{eq:shell_model})
is the Lanczos algorithm; an iterative method
which gives the solution of the lowest eigenstates. This method was 
already applied to nuclear physics problems by Whitehead {\sl et al.} 
in 1977. The technique is described in detail in \mbox{Ref. \cite{whit77}}, 
see also \mbox{Ref. \cite{ehho95}}.

\section{Results and dicussions}
The results of the shell model calculation are presented in Tables~1--4.
Our main intention is to gain insight about the effective interaction
in nuclear systems and see to what extent our calculated two--particle
matrix elements can reproduce the general features of the experimental data
in the Sn region. All experimental information in the present analysis is taken 
from the data base of the National Nuclear Data Center at \mbox{Brookhaven \cite{brook}}.

%%%%%%%%%%%%%%  table 1  %%%%%%%%%%%%%%%%%%%%
\begin{table}[t]
\caption{Exitation spectra for the light Sn isotopes. \label{tab:1}}
\vspace{0.2cm}
\begin{center}
\footnotesize
\begin{tabular}{|cccc|cccc|}
\hline \hline
&&&&&&&\\[-5pt]
\multicolumn{4}{|c|}{$^{102}$Sn}&\multicolumn{4}{|c|}{$^{104}$Sn}\\
$J^{\pi}$&Exp.&$J^{\pi}$&Theory&$J^{\pi}$&Exp.&$J^{\pi}$&Theory\\
\hline
&&&&&&&\\[-3pt]
$(2^{+})$ & $1.47$  & $2^{+}$   & $1.73$ &$(2^{+})$ & $1.26$ & $2^{+}$ & $1.42$\\
$(4^{+})$ & $1.97$  & $4^{+}$   & $2.10$ &$(4^{+})$ & $1.94$ & $4^{+}$ & $1.99$\\
$(6^{+})$ &  & $6^{+}$   & $1.96$ &$(6^{+})$ & $2.26$ & $6^{+}$ & $2.37$\\[3pt]
\hline \hline
&&&&&&&\\[-5pt]
\multicolumn{4}{|c|}{$^{106}$Sn}&\multicolumn{4}{|c|}{$^{108}$Sn}\\
$J^{\pi}$&Exp.&$J^{\pi}$&Theory&$J^{\pi}$&Exp.&$J^{\pi}$&Theory\\
\hline
&&&&&&&\\[-3pt]
$(2^{+})$ & $1.21$ & $2^{+}$   & $1.36$ &$(2^{+})$ & $1.21$ & $2^{+}$ & $1.44$\\
$(4^{+})$ & $2.02$ & $4^{+}$   & $2.15$ &$(4^{+})$ & $2.11$ & $4^{+}$ & $2.37$\\
$(6^{+})$ & $2.32$ & $6^{+}$   & $2.36$ &$(6^{+})$ & $2.37$ & $6^{+}$ & $2.47$\\[3pt]\hline
\end{tabular}
\end{center}
\end{table}
%%%%%%%%%%%%%%%%%%%  end table 1 %%%%%%%%%%%%%%

%%%%%%%%%%%%%%  table 2  %%%%%%%%%%%%%%%%%%%%
\begin{table}[t]
\caption{Exitation spectra for the heavy Sn isotopes. \label{tab:2}}
\vspace{0.2cm}
\begin{center}
\footnotesize
\begin{tabular}{|cccc|cccc|}
\hline
&&&&&&&\\[-5pt]
\multicolumn{4}{|c|}{$^{130}$Sn}&\multicolumn{4}{|c|}{$^{128}$Sn}\\
$J^{\pi}$&Exp.&$J^{\pi}$&Theory&$J^{\pi}$&Exp.&$J^{\pi}$&Theory\\
\hline
&&&&&&&\\[-3pt]
$(2^{+})$ & $1.22$ & $2^{+}$   & $1.46$ &$(2^{+})$ & $1.17$ & $2^{+}$ & $1.28$\\
$(4^{+})$ & $2.00$ & $4^{+}$   & $2.39$ &$(4^{+})$ & $2.00$ & $4^{+}$ & $2.18$\\
$(6^{+})$ & $2.26$ & $6^{+}$   & $2.64$ &$(6^{+})$ & $2.38$ & $6^{+}$ & $2.53$\\[3pt]
\hline \hline
&&&&&&&\\[-5pt]
\multicolumn{4}{|c|}{$^{126}$Sn}&\multicolumn{4}{|c|}{$^{124}$Sn}\\
$J^{\pi}$&Exp.&$J^{\pi}$&Theory&$J^{\pi}$&Exp.&$J^{\pi}$&Theory\\
\hline
&&&&&&&\\[-3pt]
$2^{+}$ & $1.14$ & $2^{+}$   & $1.21$ &$2^{+}$ & $1.13$ & $2^{+}$ & $1.17$\\
$4^{+}$ & $2.05$ & $4^{+}$   & $2.21$ &$4^{+}$ & $2.10$ & $4^{+}$ & $2.26$\\
$     $ &        & $6^{+}$   & $2.61$ &        &        & $6^{+}$ & $2.70$\\[3pt]
\hline \hline
&&&&&&&\\[-5pt]
\multicolumn{4}{|c|}{$^{122}$Sn}&\multicolumn{4}{|c|}{$^{120}$Sn}\\
$J^{\pi}$&Exp.&$J^{\pi}$&Theory&$J^{\pi}$&Exp.&$J^{\pi}$&Theory\\
\hline
&&&&&&&\\[-3pt]
$2^{+}$   & $1.14$ & $2^{+}$   & $1.15$ & $2^{+}$  & $1.17$ & $2^{+}$ & $1.14$\\
$4^{+}$   & $2.14$ & $4^{+}$   & $2.30$ & $4^{+}$  & $2.19$ & $4^{+}$ & $2.30$\\
$6^{+}$   & $2.56$ & $6^{+}$   & $2.78$ &          &        & $6^{+}$ & $2.86$\\[3pt]
\hline \hline
&&&&&&&\\[-5pt]
\multicolumn{4}{|c|}{$^{118}$Sn}&\multicolumn{4}{|c|}{$^{116}$Sn}\\
$J^{\pi}$&Exp.&$J^{\pi}$&Theory&$J^{\pi}$&Exp.&$J^{\pi}$&Theory\\
\hline
&&&&&&&\\[-3pt]
$2^{+}$   & $1.22$ & $2^{+}$   & $1.15$ & $2^{+}$  & $1.30$ & $2^{+}$ & $1.17$\\[3pt]\hline
\end{tabular}
\end{center}
\end{table}
%%%%%%%%%%%%%%%%%%%  end table 2 %%%%%%%%%%%%%%

The Sn isotopes relevant for the calculation covers the range from $^{102}$Sn to 
 $^{130}$Sn. Isotopes below  $^{116}$Sn (light Sn) are treated based on the symmetric 
$Z = N = 50$ core whereas  the isotopes above $^{116}$Sn (heavy Sn) are treated based 
on the asymmetric $Z = 50, N = 82$ core. This simplifies the shell model calculation,
but in addition it is of interest to see how successful a hole--hole effective 
interaction calculated with respect to $^{132}$Sn  is.

The results in Table~1 show excitation spectra for the light
Sn isotopes. Only some selected states are displayed. First of all, the well--known 
near constant $0^{+} - 2^{+}$ spacing is well reproduced. However the spacing
is $0.1 - 0.2$~MeV too large which indicates that our effective interaction produces  
a little too much pairing correlation  compared to the experimental data. 
We believe this is related to the interaction
between the two dominant low--lying $d_{5/2}$  and $g_{7/2}$ orbits and  the intruder orbit
$h_{11/2}$. This orbit is essential for the constant  $0^{+} - 2^{+}$ spacing throughout
the Sn isotopes due to its large degeneracy. Such intruder orbits are 
difficult to handle by our effective interaction methods and the results 
indicate that further investigation is necessary on this point. 
A more complete analysis of the excitation spectra all the way up to $^{116}$Sn
is under preparation. Preliminary analysis shows similar good agreements as in Table~1.

The resulting excitation spectra for the heavy Sn isotopes are shown in Table~2.
Again the near constant $0^{+} - 2^{+}$ spacing is well reproduced all the way down to 
$^{116}$Sn, even better than for the 
light Sn isotopes. Also the additional calculated states are in very good agreement
with experiment. However more detailed analysis of the results close to $^{116}$Sn
indicates that our effective two--particle interaction has difficulties in reproducing
the shell closure which is believed to occur in this region. The increase of the 
the $0^{+} - 2^{+}$ splitting is not as sharp as found experimentally, even if the 
phenomenon is rather weak in the case of Sn.
We have observed a similar feature around $^{48}$Ca which is generally agreed
to be a good closed shell nucleus. Here the deviation between theory and experiment 
is severe. Preliminary analysis indicates that our effective interaction
may be slightly too actractive when the two particles occupy different single--particle orbits.
This may be related to the radial wave functions which in our calculation are 
chosen to be harmonic oscillator functions.

The next set of data we have analysed is the relative binding energies. Table~3
shows the results for the light Sn isotopes. In this case data for $^{100}$Sn and 
$^{101}$Sn are not known experimentally so we have calculated binding energies relative
to $^{102}$Sn by the formula
%
%%%%%%%%%%%%%%    table 3  %%%%%%%%%%%%%%%
\begin{table}[t]
\caption{Binding energies for the light Sn isotopes relative to
$^{102}$Sn. For the definition, see Eq.\ref{eq:bind-light}}
\vspace{0.2cm}
\begin{center}
\footnotesize
\begin{tabular}{|l|r|r|r|r|r|r|}
\hline
&&&&&&\\[-3pt]
   &$^{104}$Sn &$^{106}$Sn &$^{108}$Sn &$^{110}$Sn
                              &$^{112}$Sn &$^{114}$Sn\\[5pt] \hline
&&&&&&\\[-3pt]
Experiment       &-2.45&-4.19&-4.55&-4.16&-2.77&-0.45\\
Shell Model      &-2.17&-3.99&-5.39&-6.27&-6.53&-6.28\\[3pt]
\hline
\end{tabular}
\end{center}
\end{table}
%%%%%%%%%%%%%%%%%%%  end table 3  %%%%%%%%%%%%%%%

%
\bea
BE_r[^{102 + n}Sn] =&    BE[^{102 + n}Sn] - BE[^{102 }Sn] \nonumber \\ 
                    &- n \left (BE[^{103}Sn] -  BE[^{102}Sn] \right ).
 \label{eq:bind-light}
\eea
In case of the heavy Sn isotopes the necessary data  are known and the values
in Table~4 are calculated by the formula 
\bea
BE_r[^{132 - n}Sn)] =&     BE[^{132 - n}Sn] - BE[^{132}Sn] \nonumber \\  
                    &- n  \left (BE[^{131}Sn] -  BE([^{132}Sn] \right ).
 \label{eq:bind-heavy}
\eea
%
%%%%%%%%%%%%%%%    tabel 4  %%%%%%%%%%%%%%
\begin{table}[htbp]
\caption{Binding energies for the heavy Sn isotopes.
         For the definition, see Eq.\ref{eq:bind-heavy}}
\vspace{0.2cm}
\begin{center}
\footnotesize
\begin{tabular}{|l|r|r|r|r|r|r|r|r|}
\hline
&&&&&&&&\\[-3pt]
   &$^{130}$Sn &$^{128}$Sn &$^{126}$Sn &$^{124}$Sn &$^{122}$Sn
                              &$^{120}$Sn &$^{118}$Sn &$^{116}$Sn\\[5pt] \hline
&&&&&&&&\\[-3pt]
Experiment       &-2.09&-3.64&-4.79&-5.47& -5.64& -5.26& -4.28& -2.61\\
Shell Model      &-2.24&-4.60&-6.99&-9.39&-11.77&-14.11&-16.39&-18.58\\
Mod. Shell Model &-2.09&-3.72&-4.81&-5.32& -5.22& -4.51& -3.15& -1.12\\[3pt]
\hline
\end{tabular}
\end{center}
\end{table}
%%%%%%%%%%%%%%%%%%  end table 4  %%%%%%%%%%%%%%%

For the light Sn isotopes experimental relative binding energies
show a parabola structure with a minimum around $^{108}$Sn. This is an effect of the Pauli principle
and the limited number of the degrees of freedom in the $P$--space for the valence particles.
Here the dominant orbits are $g_ {7/2}$ and $d_ {5/2}$ which should  give a minimum around
$^{108}$Sn and a shell closure around  $^{116}$Sn.
Theory produces more binding with a minimum around  $^{110}$Sn, again indicating too much 
influence of the $h_ {11/2}$ orbit.

A similar and even more dramatic result is seen in the calculation of the relative binding energies
for the heavy Sn isotopes. Experiment indicates a minimum around $^{124}$Sn--$^{122}$Sn and consequently
a shell closure around $^{116}$Sn whereas theoretical binding energies increases linearly all the way 
down to $^{116}$Sn. Thus the shell model calculation uses all $P$--space degrees of freedom to produce 
ground state binding energies in clear contradiction to experiment.   

This phenomenon of overbinding of nuclear  systems when  effective interactions from meson theory
are used have been much dicussed in the literature, see for example \mbox{Ref. \cite{andre}}.
The arguments are that such matrix elements must be modified in order to reproduce
the binding energies correctly. The so--called centroid matrix elements
should be modified in order to reproduce experiment. However, no well recipe for doing this 
is available.

In our case we have investigated the heavy Sn isotopes and defined a global centroid by
\be
W = \frac{1}{\mbox{dim}} \sum_{j_1 \ge j_2} 
                   \frac{\sum_J (2J+1) <j_1, j_2:J| V |j_1, j_2:J>}
                        {\sum_J (2J+1)}
\label{eq:cent}
\ee
where $\mbox{dim} = 160$, the total number of matrix elements in the present calculation.
In the calculation of $^{130}$S our theoretical binding energy gave $-2.24$~MeV whereas experiment
gives $-2.09$~MeV. Thus we made a global monopol correction $W \cdot (n(n-1))/2$ to all matrix elements
and adjust $W$ to reproduce the correct binding energy of $-2.09$~MeV. This gave $W = +0.15$~MeV.
The modified binding energies are displayed in Table~4, now in very good agreement with experiment.
Such a modification of the matrix elements has no effect on the excitation spectra and preserves 
the good agreement in this part of the calculation. A similar modification is not possible in the light
Sn isotopes since essential data for $^{100-102}$Sn is not available.

\section{Conclusion}
We have presented the basic elements for a calculation of a realistic microscopic
effective interation. The interaction is derived from a modern meson--exchange NN potential using
many--body perturbation theory. This is applied to the Sn isotopes ranging from
$A = 102$ to $A = 132$  where both excitation spectra and relative binding energies are calculated.
Excitation spectra are in good agreement with experiment without any adjustment of  parameters related 
to either single--particle energies or matrix elements. Relative binding energies are calculated
and show clear deviations from experiments. This indicates problems related to the methods used
in calculating effective interactions from meson theory. The BONN CD potential\mbox{potential \cite{cdbonn}}
 which is used in the present calculation produces more binding than previous versions and
may be part of the reason for overbinding in the calculation. Another possible explanation is 
related to the radial wave functions which are taken to be harmonic oscillators. This question will 
be investigated further.

We have shown that a global monopole term added to all matrix elements can cure the difficulties
with the binding energies, at least in the heavy Sn isotopes. Only a change of binding energy 
for two holes from the calculated $-2.24$~Mev to $-2.09$ is needed for good agreement with experiment.
Thus, in spite of the drastic difference shown in Table~4 we believe that only minor improvements
are necessary to give good effective interactions to be used in shell model calculations.

\section*{References}


\begin{thebibliography}{99}
%
\bibitem{hko95}  M.\ Hjorth-Jensen, T.\ T.\ S.\ Kuo and
E.\ Osnes, Phys.\ Reports 261 (1995) 125.
%
\bibitem{eh97} T.\ Engeland and M.\ Hjorth--Jensen, in preparation.
%
\bibitem{cdbonn} R.\ Machleidt, F.\ Sammarruca and Y.\ Song,
Phys.\ Rev.\ C 53 (1996)
%
\bibitem{mac89}  R.\ Machleidt, Adv.\ Nucl.\ Phys.\ 19 (1989)  189.
%
\bibitem{nim} V.G.J.\ Stoks, R.A.M.\ Klomp, C.P.F. Terheggen and J.J.\
de Swart, Phys.\ Rev.\ C 48 (1993) 792.
%
\bibitem{v18} R.B.\ Wiringa, V.G.J.\ Stoks and R.\ Schiavilla, Phys.\ Rev.\
C 51 (1995) 38.
%
\bibitem{ko90}  T.T.S.\ Kuo and E.\ Osnes, Folded-Diagram Theory of the
Effective Interaction in Atomic Nuclei, Springer Lecture Notes in Physics,
(Springer, Berlin, 1990) Vol.\ 364.
%
\bibitem{leeSu} S.Y.~Lee and K.~Suzuki, Prog. Theor.\ Phys. 64 (1980) 2091
%
\bibitem{paul} P.J.~Ellis, T.~Engeland, M.~Hjorth-Jensen, A.~Holt and E.~Osnes,
\ Nucl.\ Phys.\ A573 (1994) 216.
%
\bibitem{aldo} F.~Andreozzi, L.~Coraggio, A.~Covello, S.~Gargano, T.T.S.~Kuo and 
A.~Porrino, Phys.\ Rev.\ C 56 (1997) R16.
%
\bibitem{nico} N.~Sandulescu, A.~Blomqvist and R.J.~Liotta,
\ Nucl.\ Phys.\ A582 (1995) 216.
%
\bibitem{jan} B.~Fogelberg and J.~Blomqvist, Phys. Lett. 137B (1984) 20
%
 \bibitem{whit77} R.R.\ Whitehead, A.\ Watt, B.J.\ Cole and I.\
                  Morrison, Adv.\ Nucl.\ Phys.\ 9 (1977) 123.
%
\bibitem{ehho95}T.\ Engeland, M.\ Hjorth-Jensen, A.\ Holt and E.\ Osnes,
               Phys.\ Scripta T56 (1995) 58.
%
\bibitem{brook} National Nuclear Data Center, Brookhaven National Laboratory,
%
\bibitem{andre} A.P.~Zuker, \ Nucl.\ Phys.\ A576 (1994) 65,\\
               G.~Martinez-Pinedo, A.P.~Zuker, A.~Poves, E.~Caurier,
               Phys.\ Rev.\ C 55 (1997) 187.
%
\end{thebibliography}
\end{document}